\documentstyle{article}

\input epsf

\begin{document}

\title{Effect of long range forces on the interfacial profiles in thin binary polymer films
} 

\author{
A.\ Werner, M.\ M\"{u}ller, F.\ Schmid, and K. Binder
\\
{\small Institut f{\"u}r Physik, WA 331, Johannes Gutenberg Universit{\"a}t}
\\
{\small D-55099 Mainz, Germany}
}
\date{\today, submitted to J.Chem.Phys.}
\maketitle

\begin{abstract}
We study the effect of surface fields on the interfacial properties of a binary polymer melt confined between two parallel
walls. Each wall attracts a different component of the blend by a non-retarded van der Waals potential. An interface
which runs parallel to the surfaces is stabilized in the center of the film. Using extensive Monte Carlo simulations 
we study the interfacial properties as a function of the film thickness, the strength of the surface forces and the lateral
size over which the profiles across the film are averaged. We find evidence for capillary wave broadening of the apparent interfacial profiles.
However,
the apparent interfacial width cannot be described quantitatively by a simple logarithmic dependence on the film thickness.
The Monte Carlo simulations reveal that the surface fields give rise to an additional  reduction of the intrinsic interfacial width and an
increase of the effective interfacial tension upon decreasing the film thickness. These modifications of the intrinsic interfacial 
properties are confirmed by self-consistent field calculations. Taking account of the thickness dependence of the intrinsic interfacial
properties and the capillary wave broadening, we can describe our simulation results quantitatively.
\end{abstract}

\section{ Introduction. }
\noindent
Interface and surface properties in polymeric composites have attracted abiding interest. The surface structure of a material correlates 
with its application properties ({\em e.g.} adhesive and wetting properties, lubrification). Hence, the control of the surface and interfacial
structure is a key for tailoring novel materials.\cite{GENERAL} Surface and thin film properties also pose challenging theoretical questions. The presence of
the surface alters the phase behavior\cite{NAKANISHI,BREV} and restricts the fluctuations of an interface in the vicinity of the surface.\cite{MICHAEL} However, a direct 
comparison of surface and 
interfacial profiles between theory and experiment or computer simulation is difficult:\cite{AW} While the theoretical approaches calculate an intrinsic profile,
experimental profiles are broadened by capillary waves.\cite{KRAMER,TK1,JONESEX} These fluctuations of the local interfacial position give rise to a dependence of the width of the 
apparent interfacial profiles on the lateral resolution and the thickness of the film. Though capillary waves are present at all
fluid interfaces,\cite{CAP,JASNOV} the study of polymeric systems is particularly rewarding: The extended structure of the macromolecules facilitates the application 
of several experimental methods; {\em e.g.}, the surface excess of a component is typically an order of magnitude larger than in its low molecular 
weight counterpart. By virtue of their extension a polymer interacts with many neighbors and composition fluctuations in the bulk are small.\cite{GINZ} Hence, 
the thermodynamical properties of high molecular weight blends are well described by mean field theory and, apart from capillary wave fluctuations, self-consistent 
field theories\cite{HELFAND,FLEER,NOOLANDISCF,SHULL}  are believed to provide a detailed description of interfaces in the limit of very long chain lengths.  Hence, the effects 
of interfacial fluctuations in polymer blends can be accurately isolated and a comparison between experiments, computer simulation, and analytic theory is possible.

Many theoretical approaches\cite{NAKANISHI,BREV} assume short range interactions between the monomeric units and the surfaces. 
Experiments indicate that this is a  reasonable approximation if the range of the interactions is small compared to the polymer's extension.\cite{JONES} 
In the case of short range surface fields, the width of the apparent profile in a thin film increases like $\sqrt{D}$, where $D$ denotes the thickness 
of the film. This has been studied analytically,\cite{SOFT} experimentally\cite{TK1,TK2} and in Monte Carlo simulations.\cite{AW,BSIM,REV} 
In simple fluids, the influence of long range forces can shift the order of the wetting transition from second to first order\cite{KROLL,SCHICK} and modifies the composition 
profiles near the surface.\cite{NOOLANDI} There is also experimental evidence that long range force influence the dynamics of wetting layers in thin films.\cite{STEINER}
Spinodal dewetting in the presence of long range forces was recently investigated numerically by Puri and  Binder.\cite{PURI} Effects of long range forces have been studied
by Monte Carlo simulations by Pereira and Wang,\cite{PEREIRA} and in the framework of self-consistent field calculations by Genzer and co-workers.\cite{GENZER3} Both techniques
reveal only qualitative changes due to the presence of long range surface fields for long polymer chains. 
The presence of long range forces, however, is expected to have pronounced effects on the dependence of the interfacial width in a thin film on its thickness. In contrast to
the short range case, a much weaker, logarithmic dependence on the film thickness is predicted. This has been recently observed in experiments of Sferrazza {\em et al.}\cite{JONESEX} 

The present study aims at investigating the interplay between long range forces and the apparent interfacial profiles in a thin film geometry.
We study the dependence of the interfacial width in a binary polymer blend on the film thickness and the strength of the surface interactions via 
Monte Carlo (MC) simulations in the framework of a coarse grained lattice model. The results are compared to self-consistent field (SCF) calculations without adjustable 
parameter. This allows to separate the effect of capillary fluctuations and the influence of the surface interaction on the intrinsic interfacial profile. 
We restrict ourselves to the case of antisymmetric surface fields. {\em I.e.}, the film is confined between two walls; the right wall attracts species $A$ with the same strength and 
range as the left wall attracts species $B$. Otherwise the two walls are identical and the polymer mixture is chosen perfectly symmetric (same chain length, stiffness, {\em etc}.).

Analyzing the dependence of the interfacial width on the lateral coarse graining size\cite{AW,AW2} and measuring the spectrum of 
interfacial fluctuations\cite{MS,MW,MB} we find convincing evidence for a broadening of interfacial profiles due to capillary waves. The surface fields impart a lateral
correlation length $\xi_\|$ on the interfacial fluctuations which is determined {\em via} the direct measurement of the correlation function of composition 
fluctuations and the spectrum of interfacial fluctuations. $\xi_\|$ grows proportional to the square of the film width $D$ and, consequentially, 
the apparent interfacial width is expected to exhibit a weak logarithmic dependence $w \sim \ln D$ on the film thickness for large lateral system extensions 
$L \gg \xi_\|$. For the system sizes studied, however, our simulations
reveal an additional coupling between the long range surface fields and the intrinsic properties of the interface. The effect of the surface fields on the
intrinsic interfacial properties is confirmed by self-consistent field calculations. A quantitative analysis of our Monte Carlo results has to incorporate both effects.

Our paper is arranged as follows: The next section provides the pertinent theoretical background. In section III we describe the model that is used in the 
calculations and briefly comment on the Monte Carlo (MC) and self-consistent field (SCF) technique. Then we detail our results and close with a summary
and a discussion of our findings.

\section{Background.}
\noindent
In this section, we briefly review some of the basic theoretical predictions and provide the analytical description pertinent to the analysis of our Monte Carlo 
results. We consider a binary blend in a thin film of thickness $D$. The confining impenetrable surfaces are parallel to the $xy$ plane and located at $z=0$ and $z=D$. Each surface 
preferentially attracts a different component {\em via} a long range potential. The temperature is chosen such that the blend is phase separated in the bulk ($T<T_{cb}$) and 
the components wet their corresponding surfaces $(T>T_{\rm wet})$ in a semi-infinite slab geometry. In a thin film with antisymmetric boundaries, an interface is stabilized 
in the middle of the film which runs parallel to the surfaces. Since the interface is located on average at the center of the film the system is in the one phase region ({\em soft-mode phase}).\cite{SOFT}

Let $u(x,y)$ denote the deviations of the local interfacial position from its equilibrium position $z=D/2$. The free energy of a configuration 
can be described by an effective interfacial Hamiltonian in terms of the local interfacial position:\cite{CAPH}
\begin{equation}
\frac{{\cal H}[u]}{k_BT} = \int {\rm d}x{\rm d}y\;\; \left\{ \frac{\sigma_{\rm eff}}{2} (\nabla u)^2 + V_{\rm eff}(u) \right\}
\approx \int {\rm d}x{\rm d}y\;\; \left\{ \frac{\sigma_{\rm eff}}{2} (\nabla u)^2 + \frac{1}{2}\frac{\partial^2 V_{\rm eff}}{\partial u^2}|_{0} u^2 \right\}
\label{eqn:H}
\end{equation}
In the following we measure energies in units of $k_BT$.
The first term accounts for the free energy costs of an increase in the interfacial area caused by a fluctuating local interfacial position. For an unconfined
interface $\sigma_{\rm eff}$ is expected to equal the interfacial tension $\sigma$ between coexisting bulk phases. This has been confirmed by Monte Carlo simulations for the model employed in the calculations.\cite{MW}
The second term describes the effect of the surface fields on the  local interfacial position. For small deviations $u$ the potential can be approximated by a
parabola. Fourier decomposing the local interfacial position and using the equipartition theorem, we obtain for the mean square value of the Fourier component $u(q)$:\cite{MICHAEL}
\begin{equation}
\frac{1}{\langle |u(q)|^2 \rangle} = \sigma\left(q^2+\frac{1}{\xi_\|^2}\right)
\label{eqn:S}
\end{equation}
where the parallel correlation length for interfacial fluctuations is given by:
\begin{equation}
\xi_{\|} =  \sqrt{\frac{\sigma}{\partial^2 V_{\rm eff}(u)/\partial u^2}}
\label{eqn:X}
\end{equation}
For surface fields which decay like $z^{-n}$ the second derivative of the effective potential $V_{\rm eff}$ in the middle of the film is of the order
$D^{-n-1}$ (see below). Consequently, the parallel correlation length increases proportional to $D^{+(n+1)/2}$. This length scale acts as a cut-off for the 
spectrum of interfacial fluctuations 
and limits the broadening of the apparent profile. The fluctuations of the local interfacial position $u(x,y)$ are Gaussian distributed and its variance is obtained by integrating
over all lateral Fourier components. The power law spectrum of the free interface $(V_{\rm eff}=0)$ leads to logarithmic divergences for $q \to 0$ and $q \to \infty$. We 
can remove these divergences by introducing heuristic cut-offs $q_{\rm max}$ and $q_{\rm min}$, and obtain
\begin{equation}
\langle u^2(x,y) \rangle = \frac{1}{4\pi\sigma} \ln \left(\frac{q_{\rm max}^2 + \xi_\|^{-2}}{q_{\rm min}^2 + \xi_\|^{-2}} \right)
\end{equation}
In a system of finite lateral extension $L$ the lower cut-off is simply determined by the system size $q_{\rm min} = 2\pi/L$. The value of the
upper cut-off can be {\em defined} as the lateral size $B_0$ on which the width of the apparent interfacial profile coincides with the width of the intrinsic  profile
calculated in the self-consistent field calculations.\cite{AW2}

We approximate the laterally averaged apparent profile by the convolution of the intrinsic profile $\rho_{0}(z)$ with the Gaussian distribution
of the local interfacial position $u(x,y)$. This approximation is appropriate for an unconfined interface, but it neglects a possible dependence of the
intrinsic profile on the local interfacial position, {\em i.e.}, the distance between the interface and the wall.
Defining the width of the interfacial profile by the inverse slope at the center, this convolution approximation yields:\cite{AW,JASNOV,SEM,GREST}
\begin{equation}
w^2 = w_0^2 + \frac{2 w_0^3}{\pi} \left| \frac{{\rm d}^3\rho_0}{{\rm d}z^3}\right|_0 \langle u^2 \rangle \approx w_0^2 + \frac{1}{8\sigma} \ln \left(\frac{(2\pi/B_0)^2 + \xi_\|^{-2}}{(2\pi/L)^2 + \xi_\|^{-2}} \right)
\label{eqn:W}
\end{equation}
where we have approximated the shape of the intrinsic profile by an error function in the last step in order to calculate the numerical prefactor.

In the limit $B_0 \ll \xi_{\|} \sim D^2 \ll L$ the apparent interfacial width increases like 
$w^2 \sim (1/2\sigma) \ln D$
with the film thickness for non-retarded van der Waals interactions ($n=3$). Sferrazza {\em et al.} found experimental evidence for such a weak logarithmic dependence 
of the apparent width on the film thickness in PS/PMMA polymer blends.\cite{JONESEX} However the interfacial tension $\sigma$ extracted from the thickness dependence was 
about a factor $1.6$
larger than estimated by other means. The aim of our Monte Carlo simulations is to investigate the influence of long range surface fields on the interfacial width 
taking advantage of the simultaneous access to the interfacial structure on various lateral length scales in the Monte Carlo simulations.

\section{Model and computational techniques.}
\noindent
Coarse grained models of polymeric systems have proven useful for investigating universal properties of polymer blends and solutions.
We employ the bond fluctuation model\cite{BFM} on a three dimensional simple cubic lattice. This model retains the relevant features of polymeric materials --
connectivity of the monomers along a chain, excluded volume interaction of the segments, and a short range thermal interaction potential between monomer units --
and is highly computationally efficient. In the framework of the bond fluctuation model each segment occupies  all eight corners of a unit cell, and no site can be 
doubly occupied. Adjacent monomers along a chain molecule are connected {\em via} one of 108 bond vectors of length $2,\sqrt{5},\sqrt{6},3$, or $\sqrt{10}$.
Here and in the following all lengths are measured in units of the lattice spacing. We work at a filling fraction $8\Phi=0.5$ which corresponds to a melt or a concentrated solution.\cite{PAUL}
The binary blend consists of two species denoted $A$ and $B$. Both species are structurally symmetric and comprise $N=32$ monomeric units. Since each monomer in 
our coarse grained model corresponds to 3-5 chemical repeat units the chain length corresponds to a degree of polymerization of the order 100.

The monomers interact {\em via} a square well potential which is extended over all 54 neighbors sites up to a distance $\sqrt{6}$. This includes all lattice sites inside
the first peak of the density-density pair correlations function. A pair of monomers of the same type in the range of the square well potential lowers the energy by
$\epsilon$, whereas unlike species increase the energy by the same amount. We choose the well depth $\epsilon=0.03 $, a value which corresponds to a rather strong
segregation $\chi N \approx 5.1$ where $\chi$ denotes the Flory Huggins parameter.\cite{MM}
Many properties of this coarse grained model have been determined in prior studies:
The radius of gyration is $R_g=6.96$. The interfacial tension $\sigma=0.015 $ has been measured {\em via}  reweighting techniques\cite{MBO} and it gratifyingly agrees with the value measured 
from the spectrum of interfacial fluctuations of a free interface.\cite{MW} The correlation length $\xi_b$ of composition fluctuations\cite{AW} in the bulk is 3.8 and the intrinsic width of the free interface
is $4.6$ in self-consistent field calculations.\cite{MW,SM}

\begin{figure}[htbp]
    \begin{minipage}[t]{160mm}%
       \mbox{
           \setlength{\epsfxsize}{9cm}
           \epsffile{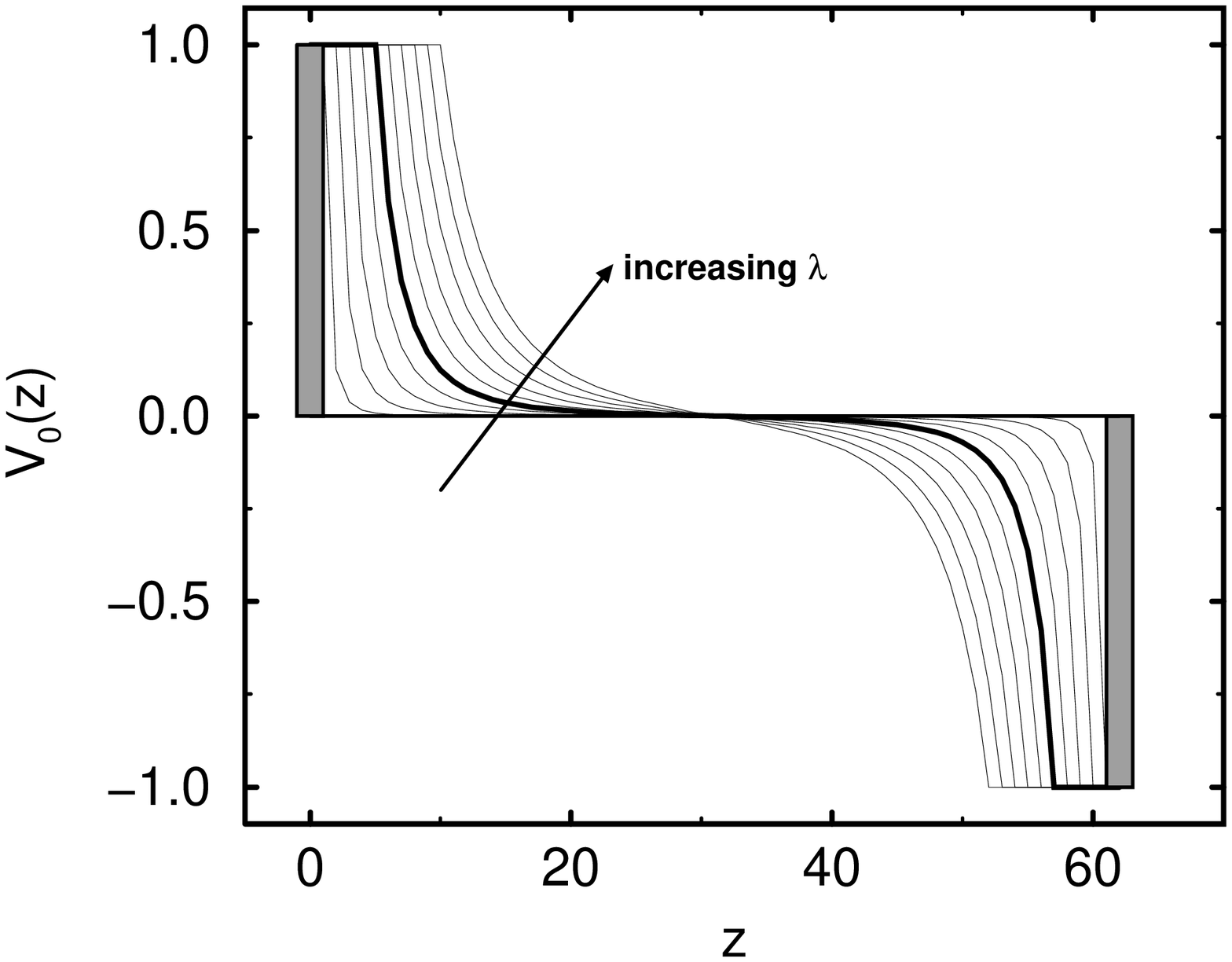}
       }
    \end{minipage}%
    \hfill%
    \begin{minipage}[b]{160mm}%
       \caption{Illustration of the long range surface field $V_0(z)$ acting on an $A$ monomer at position $z$.
                The parameter $\lambda$ tunes the shape of the surface interactions ($1 \leq \lambda \leq 10$). The shaded region corresponds to purely short range surface interactions.
		The thick curves marks the potential for $\lambda=5$.
                }
       \label{fig:pot}
    \end{minipage}%
\end{figure}

The confining surfaces are modelled as ideally flat, parallel, and structureless. The left wall attracts the
$A$ component whereas the right one the $B$ species. The long range potential between the monomers and the wall is taken to be  $\pm \epsilon_w V_0(z)$, where
the positive sign holds for $A$ monomers and the negative for $B$ ones. The shape of the potential is given by
\begin{equation}
V_0(z) = \left\{                                   
       \begin{array}{lll}
       1                                                     & \mbox{for} & z \leq \lambda \\
       \frac{\lambda^3}{z^3} - \frac{\lambda^3}{(D-2-z)^3}   & \mbox{for} & \lambda < z < D-2-\lambda  \\
       -1                                                    & \mbox{for} &  D-2-\lambda \leq z
       \end{array}
       \right.
\end{equation}
as to mimic non-retarded van der Waals interactions between the monomers and the surface molecules. The parameter combination $\epsilon_w \lambda^3$ corresponds to the Hamaker constant. 
The monomer coordinates refer to the left corner of the monomers, hence the $z$ coordinates range
of a monomer range in the interval [0,D-2]. The left wall attracts $B$ monomers at position $z$ as much as the right wall attracts $A$ ones 
at position $D-2-z$, {\em i.e.}, the surface fields are completely antisymmetric. This potential shape is illustrated in Fig.\ref{fig:pot}.
The strength of the potential is held constant at $\epsilon_w=0.1 $, whereas we vary the range $1 \leq \lambda \leq 15$.
Prior studies\cite{AW,MB} show that at this combination of parameters the $A$ phase wets the surface in a semi-infinite geometry for purely short range surface fields
which are extended over the two nearest layers to the wall. Since the long range surface fields are always stronger, the system is above the wetting 
temperatures for all values of $\lambda$. In a thin film geometry with antisymmetric surface fields an interface is stabilized which runs parallel to 
the walls and is located on average in the center of the film.\cite{SOFT}

Simulations are performed in the canonical ensemble and in the semi-grandcanonical ensemble.
To relax the polymer conformations on the lattice we use a combination of random local monomer displacements and slithering snake moves.\cite{MM}
While the  composition of the system is fixed at ${\rho}_A=0.5$ in the canonical runs, the composition fluctuates in the semi-grandcanonical
simulations and the exchange chemical potential $\Delta \mu$ between the polymer species is controlled. By virtue of the symmetry with respect to exchanging 
the monomer labels $A \rightleftharpoons B$ and the direction of the $z$ axis, the average interfacial position fluctuates around the center of the film at 
$\Delta \mu =0 $. Though all other than local hopping moves do not correspond to a physically realistic dynamics they yield a reasonably fast equilibration.
The different Monte Carlo moves are applied in the ratio: local hopping : slithering snake : semi-grandcanonical moves = 4:12:0.4
as to equilibrate the polymer conformations and the composition fluctuations on roughly the same time scale.

To investigate the interfacial structure on different lateral length scales (block analysis),\cite{AW} we divide the system laterally ({\em i.e.} parallel to the interface) into subsystems of size $B \times B$.
In each of these $B \times B \times D$ columns we determine the local interfacial position $u(x,y)$ using the concept of the Gibbs dividing surface.\cite{AW}
Then we average the profiles with respect to this position and determine its width. Employing this local position of the interface we calculate
the spectrum of interfacial fluctuations.\cite{MS,MW}

In order to distinguish between the influence of the long range surface fields on the capillary wave broadening and the intrinsic interfacial properties
we investigate the same model in the framework of the self-consistent (SCF) field theory.  A detailed account of our self-consistent field (SCF) method has been given elsewhere\cite{MW,MB}
and we only summarize the salient features pertinent to the present study. In the mean field approximation the {\em inter}molecular interactions are replaced by potentials
of mean field. The potential $w_A$ acting on the $i$th monomer of $A$ polymer $\alpha$ at position $r_{\alpha,i}$ takes the form
\begin{equation}
w_A(r_{\alpha,i}) = - {\tt z}\epsilon \Big(1+ \frac{1}{2}l_0^2\partial_z^2 \Big)\left[\rho_A(r_{\alpha,i})-\rho_B(r_{\alpha,i})\right] +  \zeta\left[\rho_A(r_{\alpha,i})+\rho_B(r_{\alpha,i})-1\right]
\end{equation}
and a similar expression holds for $w_B$. The first term describes the intermolecular interaction of an $A$ monomer {\em via} the square well potential of depth $\epsilon$
with its surrounding. ${\tt z}=2.65$ denotes the number of monomers of other chains in the interaction range.\cite{MM} It has been extracted from MC simulations of the intermolecular 
paircorrelation function of a dense melt. The derivative in $z$ direction takes account of the finite range $l_0=16/9$\cite{MB,FREDDI} of the square well potential. The expression mimics 
the ``missing neighbor'' effect at the surface.\cite{HELFAND} The second term represents the free energy cost of deviations of the local monomer density $\rho_A+\rho_B$ 
from the reference value 1. This free volume contribution is approximated by an expression introduced by Helfand,\cite{HELFAND} and the compressibility $\zeta=4.1$\cite{CP} has been 
determined from Monte Carlo simulations of the athermal ($\epsilon=0$) system.

The local fields depend on the local densities which, in turn, are calculated by the average of single chain conformations in the potential of mean field and the surface field:
\begin{equation}
\rho_A(r) = \bar{\rho}_A \frac{   \sum_{\alpha=1}^{C} {\cal P}_w[r]\frac{1}{N} \sum_{i=1}^{N_A} V \delta(r-r_{\alpha,i})
                               \exp \left( -\sum_{i=1}^{N} w_A(r_{\alpha,i})  \right)                       }
                               {   \sum_{\alpha=1}^{C} {\cal P}_w[r]
                               \exp \left( -\sum_{i=1}^{N} w_A(r_{\alpha,i})  \right)                       }
\end{equation}
${\cal P}_w$ is the Boltzmann weight of an isolated chain with respect to the surface fields, {\em i.e.}, ${\cal P}_w$ vanishes if the $z$ coordinate of a 
segment is located outside the interval $[0,D-2]$ and otherwise takes the form $\exp(\sum_{i=1}^N \epsilon_w V_0(z_i))$. The summation runs over a representative
sample of up to 15 million single chain conformations which have been extracted from bulk simulations of the bond fluctuation model at the same density. We evaluate this average
on a CRAY T3E parallel computer, assigning a subset of conformations to each processor. The calculations incorporate the detailed chain structure on all length 
scales without any adjustable parameter. The above equations are expanded into a Fourier series\cite{FOURIER} and solved self-consistently.
We compare our simulational results quantitatively to the results of the self-consistent field calculations. The self-consistent field scheme describes the detailed structure of 
the unconfined interface quantitatively,\cite{MW} except for the broadening due to capillary waves. It has 
also been successfully applied to studying wetting and capillary condensation.\cite{MB}

\section{Results.}
\begin{figure}[htbp]
    \begin{minipage}[t]{160mm}%
       \mbox{
           \hspace*{-2cm}
           \setlength{\epsfxsize}{9cm}
           \epsffile{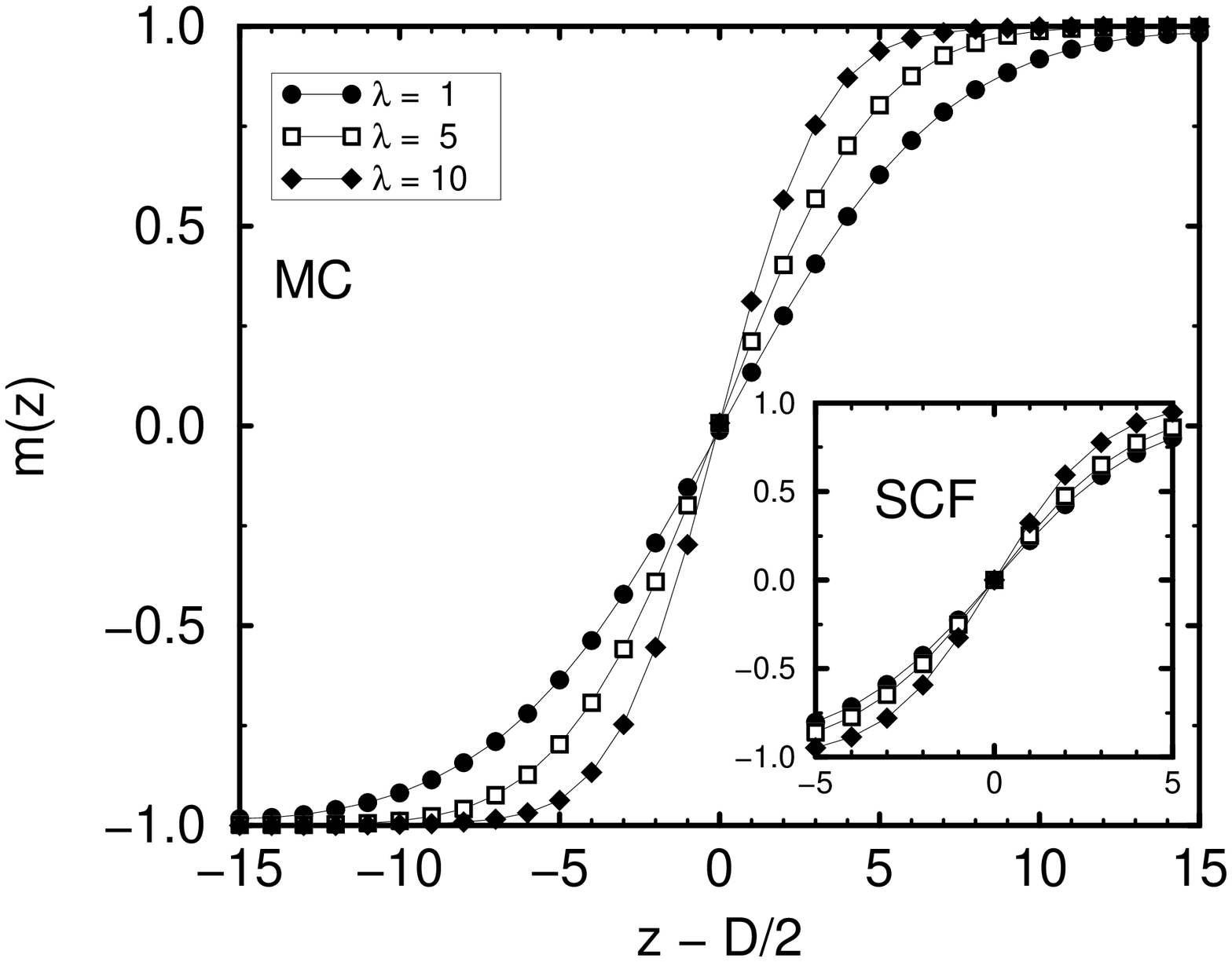}
           \setlength{\epsfxsize}{9cm}
           \epsffile{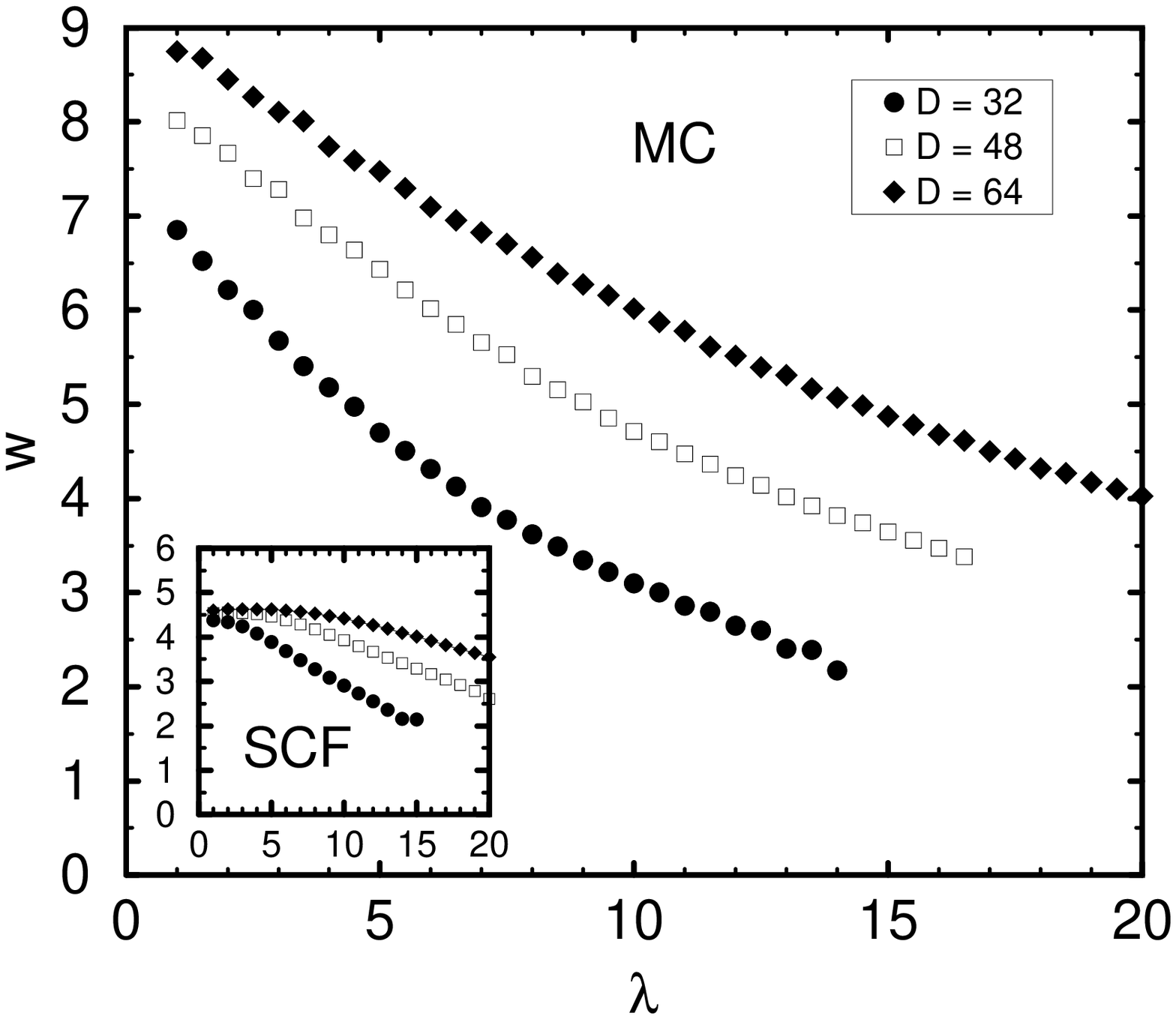}
       }
    \end{minipage}%
    \hfill%
    \begin{minipage}[b]{160mm}%
       \caption{({\bf a}) Apparent interfacial profile in the Monte Carlo simulations ($L=256$) and intrinsic profiles in the self-consistent field calculations (inset)
                for $D=32$, $\epsilon_w=0.1$, and $\epsilon=0.03$. The different curves refer to $\lambda=1,5$, and $10$. 
		 ({\bf b}) Dependence of the apparent interfacial width on the shape parameter $\lambda$ for $D=32,48$ and $64$ and lateral system size $L=256$ in the Monte Carlo
		simulations. The inset displays the result of the self-consistent field calculations for the same parameters.   
                }
       \label{fig:profile}
    \end{minipage}%
\end{figure}

\noindent
We illustrate the influence of the surface fields and the film thickness on the interfacial profiles in  Fig.\ref{fig:profile}({\bf a}). It displays the apparent profile of the 
order parameter $m(z)=[\rho_A(z)-\rho_B(z)]/[\rho_A(z)+\rho_B(z)]$, {\em i.e.}, the profile is averaged over the complete lateral extension $L=256$ of the simulation cell.
Increasing the strength of the surface fields $\lambda$ at constant film thickness results in a decrease of the width of the apparent interfacial profile for lateral extension $L=256$
by a factor of 2. The inset presents the results of the SCF calculations for the same parameters. The interfacial profiles in the SCF calculations are narrower, 
indicating the broadening of the apparent profiles by capillary waves. The SCF data also exhibit a dependence on the potential parameter $\lambda$ which is, however, weaker than in the MC
simulations. Fitting the order parameter profile to the expression $m(z) = {\rm tanh}[(z-D/2)/w]$, we determine its width $w$. The dependence of the apparent interfacial width on the film thickness $D$ and the potential shape $\lambda$
is presented in Fig.\ref{fig:profile}({\bf b}). Increasing $\lambda$ we reduce the interfacial width. A similar, though weaker dependence is found in the SCF calculations. Moreover
the apparent width in the MC simulations increases stronger than in the SCF calculations upon increasing the film thickness at small and moderate values of $\lambda$. This additional
dependence in the simulations can be attributed to  capillary wave broadening: the larger the film thickness the larger is the lateral correlation length and the stronger is broadening due to fluctuations
of the local interfacial position.

\begin{figure}[htbp]
    \begin{minipage}[t]{160mm}%
       \mbox{
           \hspace*{-2cm}
           \setlength{\epsfxsize}{9cm}
           \epsffile{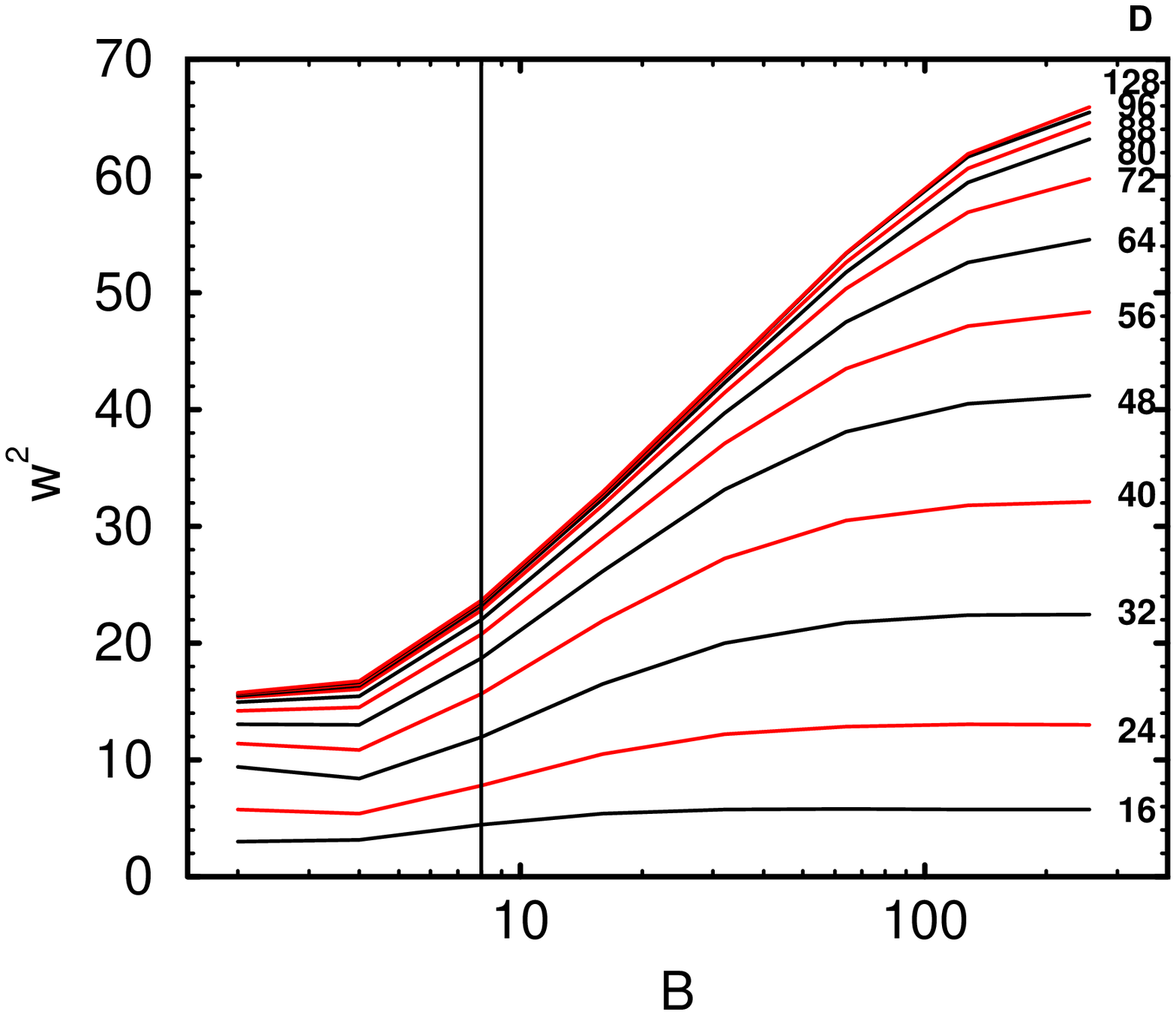}
           \setlength{\epsfxsize}{9cm}
           \epsffile{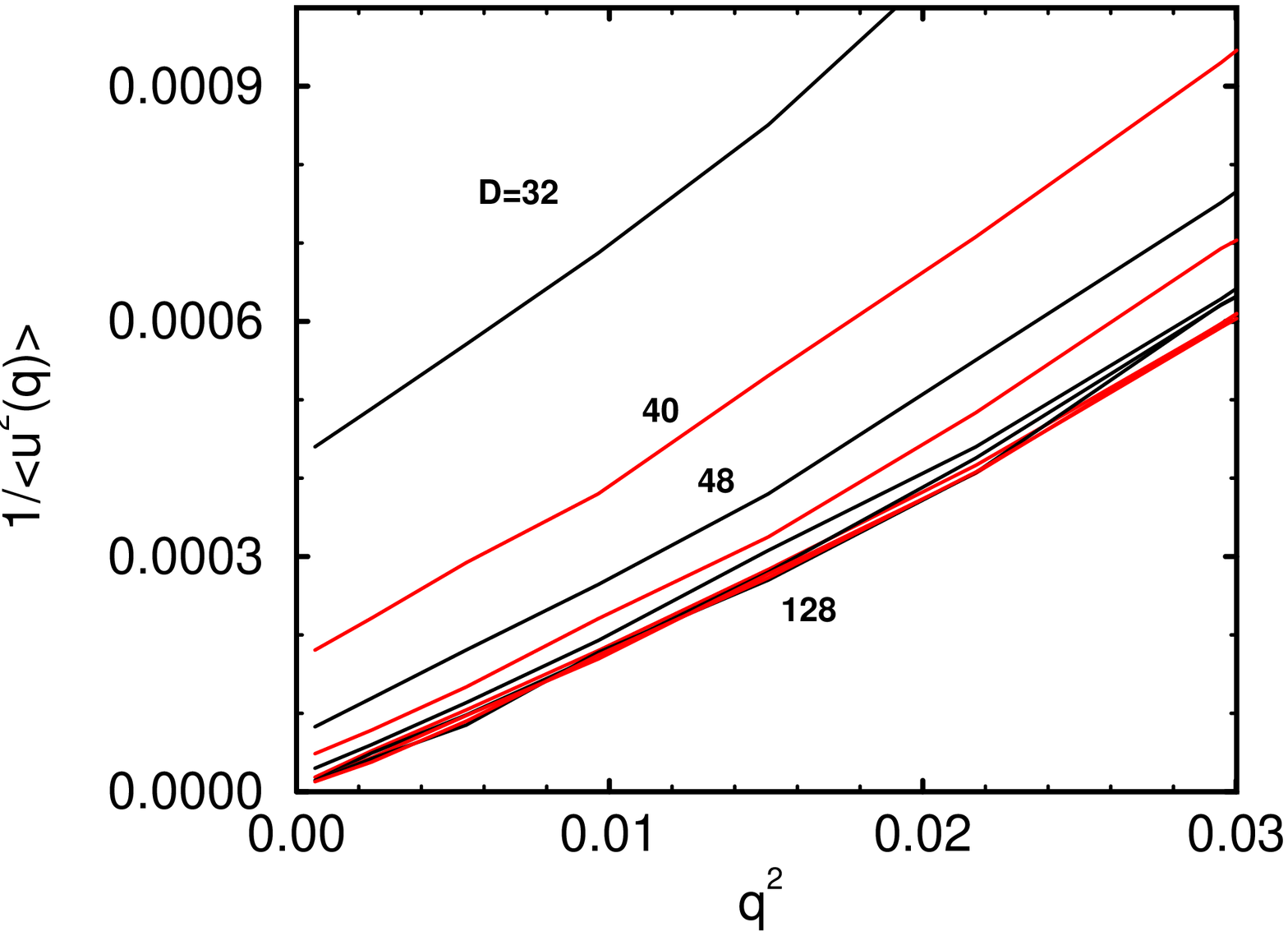}
       }
    \end{minipage}%
    \hfill%
    \begin{minipage}[b]{160mm}%
       \caption{Influence of the lateral coarse graining size on the interfacial properties:
                ({\bf a}) Dependence of the interfacial width on the lateral block size $B$ for $\lambda=5$ and different film thicknesses $D$.
                ({\bf b}) Fourier spectrum of interfacial fluctuations for $\lambda=5$ and different film thicknesses $D$.
                }
       \label{fig:block}
    \end{minipage}%
\end{figure}

To proceed in quantifying the effect of capillary waves, we investigate the dependence of the interfacial width on the lateral coarse graining size,\cite{AW} which is set by the length $B$ of the
columns used in the averaging procedure. The MC results for $\lambda=5$ are shown in Fig.\ref{fig:block}({\bf a}) for several film thicknesses $D$. For the larger film thickness, the interfacial width
increases logarithmically with the block size. Consequentially, the upper cut-off for the capillary wave broadening is set by the lateral coarse graining size $q_{\rm min} = 2\pi/B$ rather than by 
the thickness dependent parallel correlation length $\xi_{\|}$. Upon decreasing the film thickness, there are deviations from the behavior of a free interface; the interfacial width at large
block size $B$ tends to a limiting value which depends on the film thickness. This shows the limiting of the broadening due to the parallel correlation length, which is smaller than the
lateral block size $B$. For film thickness which are comparable to the polymer's end-to-end distance there is almost no influence of the block size on the interfacial width. 

The vertical line in Fig.\ref{fig:profile}({\bf a}) marks the lateral length scale $B_0=8$ on which the apparent interfacial width of a free interface\cite{AW} coincides approximately with the results 
of the SCF calculation 
$w_{\rm SCF} = 4.6$. This yields an estimate for the short wavelength cut-off, beyond which fluctuations are describable {\em via} the intrinsic profile. However,
even on this length scale the interfacial width decreases upon narrowing the film thickness. An effect of similar magnitude is also observed in the SCF calculations (cf.\
Fig.\ref{fig:profile}({\bf b})). Hence the surface fields modify the intrinsic properties of the interface. A possible dependence of the short wavelength cut-off on the film 
thickness appears to be only a minor effect because of the agreement between MC and SCF results.

In Fig.\ref{fig:block}({\bf b}) we complement this real-space analysis of the apparent interfacial width with the measurement of the Fourier spectrum of interfacial fluctuations.\cite{MS,MW} 
In accord with the
theoretical expectation the inverse mean square amplitude of the Fourier components of the local interfacial position depends quadratically on the lateral wavevector $q$. The slope of the
curve yields the interfacial tension $\sigma_{\rm eff}$ of the capillary wave Hamiltonian,\cite{MW,MB} while the abscissa of the line gives an estimate for the inverse of the lateral correlation length
(cf.\ eqn.(\ref{eqn:S})). Upon increasing the film thickness we increase the lateral correlation length and the effective interfacial tension approaches the interfacial tension of a free interface 
from above.

\begin{figure}[htbp]
    \begin{minipage}[t]{160mm}%
       \mbox{
           \setlength{\epsfxsize}{9cm}
           \epsffile{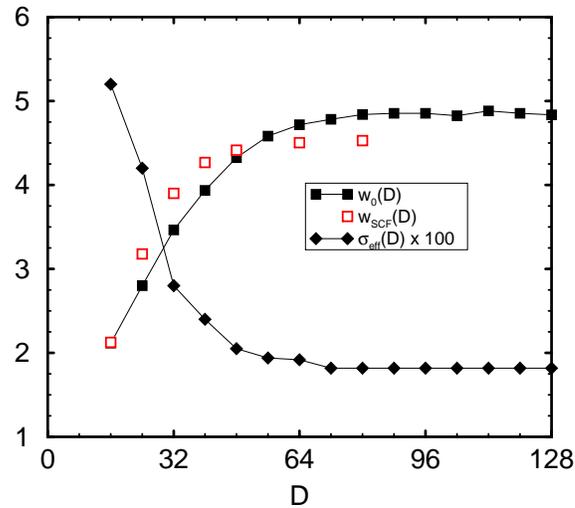}
       }
    \end{minipage}%
    \hfill%
    \begin{minipage}[b]{160mm}%
       \caption{Dependence of the interfacial width at lateral block size $B=8$ and the effective interfacial tension as extracted from the interfacial
                fluctuation spectrum  on the film thickness $D$ at $\lambda=5$. The open symbols present the result for the interfacial width in the self-consistent field calculations.
                }
       \label{fig:intrinsic}
    \end{minipage}%
\end{figure}

The results of the intrinsic properties for the interfacial width $w(B_0=8,D)$ and the effective interfacial tension $\sigma_{\rm eff}(D)$ are summarized in Fig.\ref{fig:intrinsic}.
Reducing the film thickness to about $2R_g$, we decrease the intrinsic interfacial width by more than a factor 2. This result is in qualitative agreement with the SCF calculations. The
small quantitative discrepancies between the MC result at fixed block size and the SCF calculations are probably due to residual effects of capillary waves and difficulties of the
SCF calculations to capture quantitatively the detailed structure ({\em e.g.}, packing effects) of the polymeric fluid in the vicinity of the surface.\cite{MB} With the decrease of the intrinsic interfacial width goes 
along a pronounced increase of the effective interfacial tension by more than a factor 2.5. Increasing the potential parameter $\lambda$ has similar effects on the intrinsic properties as
 reducing the film thickness $D$.

\begin{figure}[htbp]
    \begin{minipage}[t]{160mm}%
       \mbox{
           \hspace*{-2cm}
           \setlength{\epsfxsize}{9cm}
           \epsffile{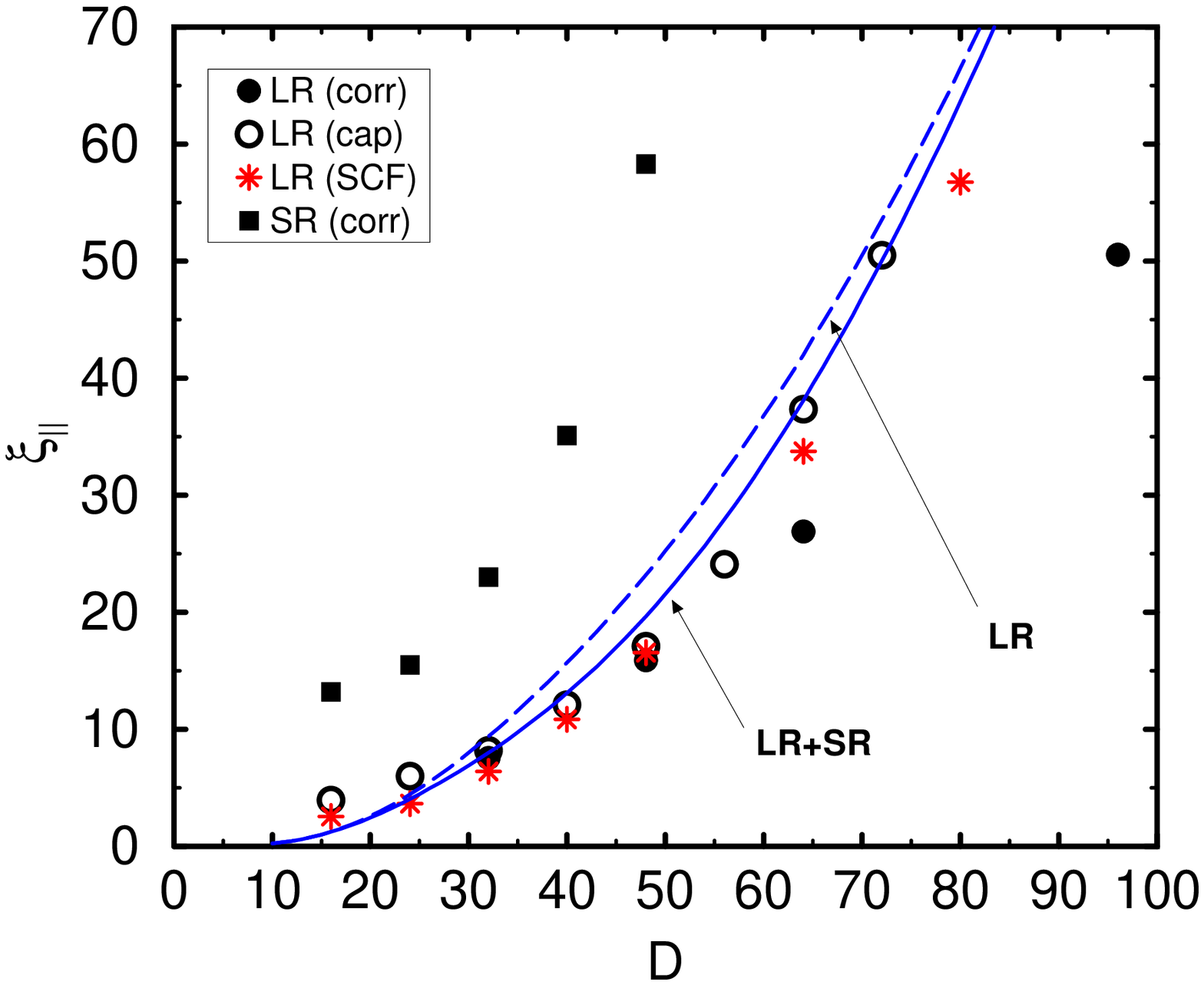}
           \setlength{\epsfxsize}{9cm}
           \epsffile{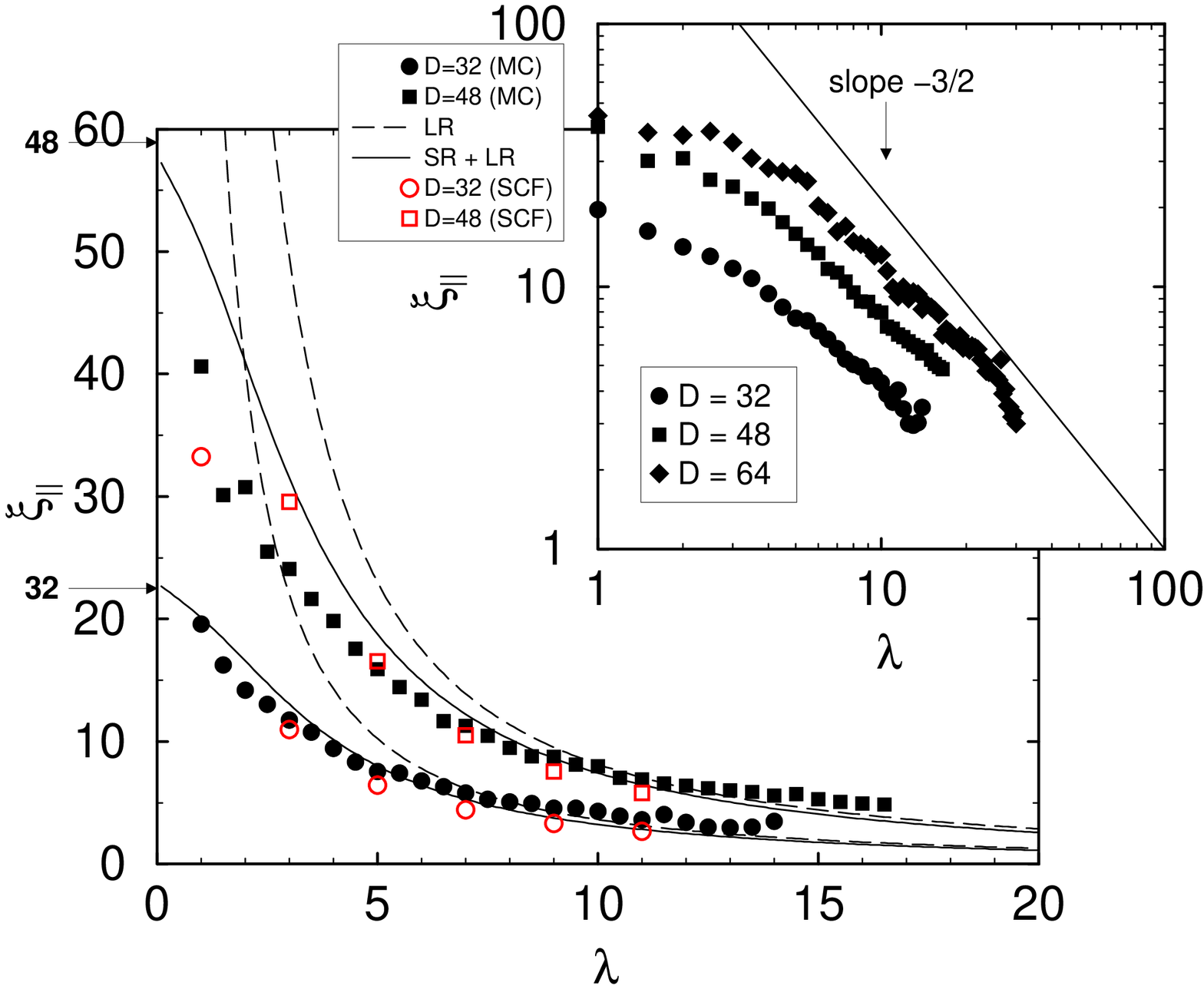}
       }
    \end{minipage}%
    \hfill%
    \begin{minipage}[b]{160mm}%
       \caption{
       \label{fig:corr}
       }
                ({\bf a}) 
		Parallel correlation length as a function of the film thickness $D$ for $\lambda=5$. Filled squares correspond to the measurement {\em via} the
		correlation function $g(r,z=D/2)$, whereas open circles present the analysis of the interfacial fluctuation spectrum. Stars denote the results of the self-consistent
		field calculations. The Monte Carlo results for purely short range surface fields\cite{AW} are depicted as filled squares. The dashed line corresponds to purely long range surface fields,
		the solid one presents the results of combined short range and long range effects according to eqns (\ref{eqn:XL}) and (\ref{eqn:XC}). 
		({\bf b}) 
                Dependence of the parallel correlation length on the potential parameter $\lambda$ for $D=32$ and $48$. Filled symbols correspond to Monte Carlo data,
                whereas open ones denote the results of the self-consistent field calculations. Dashed and solid curves correspond to purely long range forces and the combination of
		short range and long range forces, respectively. The arrows on the left hand side indicate the values for
		purely short range surface fields. The inset displays the Monte Carlo data on a log-log scale. 
    \end{minipage}%
\end{figure}

The thickness dependence of the lateral correlation length is presented in Fig.\ref{fig:corr}({\bf a}). It can be rationalized as follows.
The energy of the interface placed into the long range surface field can be described by an effective potential $V_{\rm LR}(u)$.
Assuming that the interfacial profile of the interface centered at $z=D/2+u$ takes the form $\rho_A(z) = 1- \rho_B(z) = (1+{\rm tanh}[u/w_0]/2$, 
we can calculate the long range contribution to the energy:
\begin{eqnarray}
V_{\rm LR}(u) &\approx& \int_0^D {\rm d}z\; \Phi \epsilon_w V_0(z) \left[ \rho_A(z) - \rho_B(z) \right] \nonumber \\
\frac{\partial^2 V_{\rm LR}(u)}{\partial u^2} &\approx& \frac{192 \Phi \epsilon_w \lambda^3}{D^4} 
              \left( 1 + \frac{10\pi^2}{3} \left(\frac{w_0}{D}\right)^2 + 1591  \left(\frac{w_0}{D}\right)^4 + \cdots \right)
\end{eqnarray}
where $\Phi$ denotes the monomer number density.
The leading term represents the energy of a kink-like interface in the surface field. To this leading order, the parallel correlation length takes the form
\begin{equation}
\xi_{\|{\rm LR}} \approx \sqrt{\frac{\sigma D^4}{192\Phi\epsilon_w\lambda^3}} \sim D^2
\label{eqn:XL}
\end{equation}
As mentioned above, the correlation length increases with the square of the film thickness $D$.

In addition to the corrections due to the finite width $w_0$ of the intrinsic profile, the surface fields alter the composition profile in the vicinity of the wall.
This distortion decays exponentially with the distance from the wall and  gives rise to a short range interaction between the surface and the interface.
The effect of this short range contribution can be satisfactorily described by a parallel correlation length $\xi_{\|{\rm SR}}$ of the form\cite{AW}
\begin{equation}
\xi_{\|{\rm SR}} \approx  \kappa^{-1}\exp\left(\frac{\kappa(D-2\lambda)}{4} \right)
\label{eqn:XS}
\end{equation}
where $\xi_b$ denotes the correlation length of composition fluctuations in the bulk and $\kappa^{-1}=\xi_b(1+1/8\pi\xi_b^2\sigma)$ the range of the effective interaction between the surface and the interface.
In the absence of long range surface fields the parallel correlation length increases exponentially with the film thickness $D$. 
In the present case we {\em assume} that the effective interactions of the surfaces with the interface are additive and the inverse squares of
both parallel correlation lengths add, respectively:
\begin{equation}
\xi_\|^{-2} \approx \xi_{\|{\rm SR}}^{-2}+\xi_{\|{\rm LR}}^{-2}
\label{eqn:XC}
\end{equation}
In Fig.\ref{fig:corr} ({\bf a}) we compare this approximation with the MC results and SCF calculations. In the simulations the parallel correlations length has been determined
{\em via} the spectrum of interfacial fluctuations and the direct measurement of the composition correlation functions\cite{AW}
\begin{equation}
g(r,z) \equiv \frac{\langle m(r,z) m(0,z) \rangle - \langle m(0,z) \rangle^2 }{ \langle  m(0,z)^2 \rangle - \langle m(0,z) \rangle^2 }
\end{equation}
In the middle of the film we can estimate $\xi_\|$ {\em via} $g(r,z=D/2) \sim 1/\sqrt{r} \exp(-r/\xi_\|)$. The results of both independent measurements
agree quantitatively for small $D$, whereas the differences at larger $D$ are due to statistical errors. The figure shows that the long range forces alone yield 
a too large estimate for the parallel correlation length $\xi_\|$, whereas the combination of long range and short range surface fields gives a reasonable quantitative
description of the simulation data. This is also confirmed by detailed SCF calculations. In the SCF framework we measure the free energy of a film as a function
of its composition $\rho=1/2+u/D$ around $\rho=1/2$ and employ the bulk value of the interfacial tension $\sigma$ to compute the lateral correlation length according to eqn.(\ref{eqn:X}).
For comparison we also present the lateral correlation length in the case of purely short range surface. For the parameter combination investigated, $\xi_{\rm SR}$ is always much larger
than its long range counterpart -- a fact which shows that the behavior is dominated by the long range surface fields. 

\begin{figure}[htbp]
    \begin{minipage}[t]{160mm}%
       \mbox{
           \setlength{\epsfxsize}{9cm}
          \epsffile{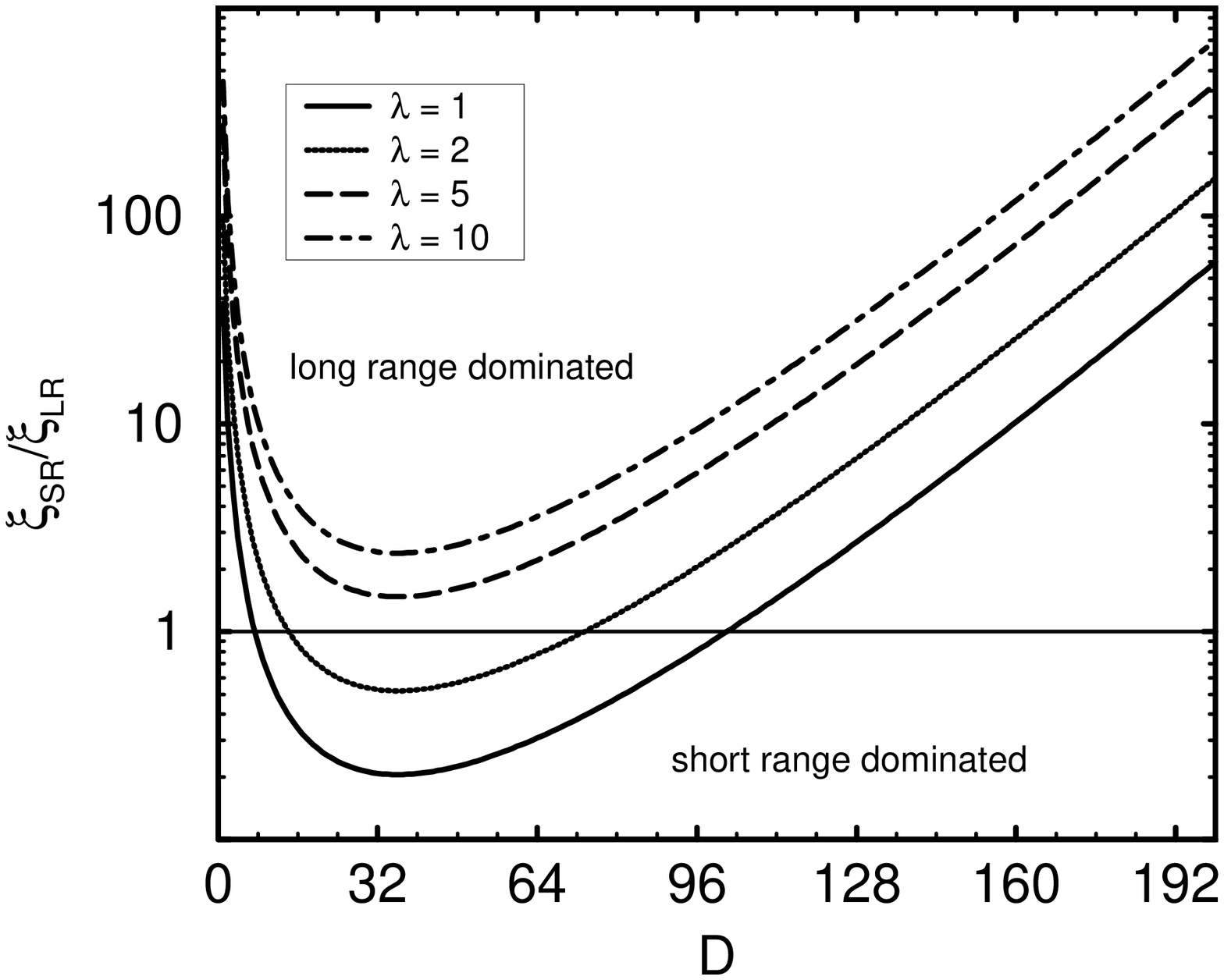}
       }
    \end{minipage}%
    \hfill%
    \begin{minipage}[b]{160mm}%
       \caption{
       \label{fig:ratio}
       }
                Ratio of the correlation length for short range surface fields and long range surface fields according to eqns.(\ref{eqn:XS}) and (\ref{eqn:XL}) as a function
	        of the film thickness and the strength of the surface fields $\lambda$. For large film thicknesses $D$ the long range forces always dominate, whereas the effect of 
		short range forces becomes appreciable for small $D\approx 5R_g$ and small values of $\lambda$.
    \end{minipage}%
\end{figure}

In Fig.\ref{fig:corr}({\bf b}) we display the dependence of the parallel 
correlation length on the surface field $\lambda$ for thickness $D=32$ and $48$. For small surface fields $\lambda \approx 1$, the correlation length approaches the value expected for
short range forces, while for larger strength of the surface fields the long range contribution dominates, and we expect $\xi_\| \sim \lambda^{-3/2}$. The limiting power law behavior is, 
however, only reached for fairly large surface fields (cf.\ inset). The analytic approximation above describes the crossover between the short range dominated ($\lambda$ small) to the long range dominated
($\lambda$ large) reasonable well and the behavior is also reproduced by our SCF calculations. The SCF calculations agree very well for large and moderate values of $\lambda$, whereas there are deviations
for small $\lambda$. In the latter case the short range contribution dominates and the lateral correlation length stems from the truncation of the interfacial profile at the wall. The deviations
between the MC results and the SCF calculations might indicate that the SCF calculations capture  only qualitatively the fine details of the profile at the wall.
We present in Fig.\ref{fig:ratio} the ratio between the purely long range 
(cf.\ eqn.(\ref{eqn:XL})) and short range contributions (cf.\ eqn.(\ref{eqn:XS})). This illustration shows that the long range forces are always dominant for large film 
thickness $D$ whereas the regime where short range surface fields dominate the behavior is restricted to rather small values of $\lambda$ and film thickness $D$.

\begin{figure}[htbp]
    \begin{minipage}[t]{160mm}%
       \mbox{
           \hspace*{-2cm}
           \setlength{\epsfxsize}{9cm}
           \epsffile{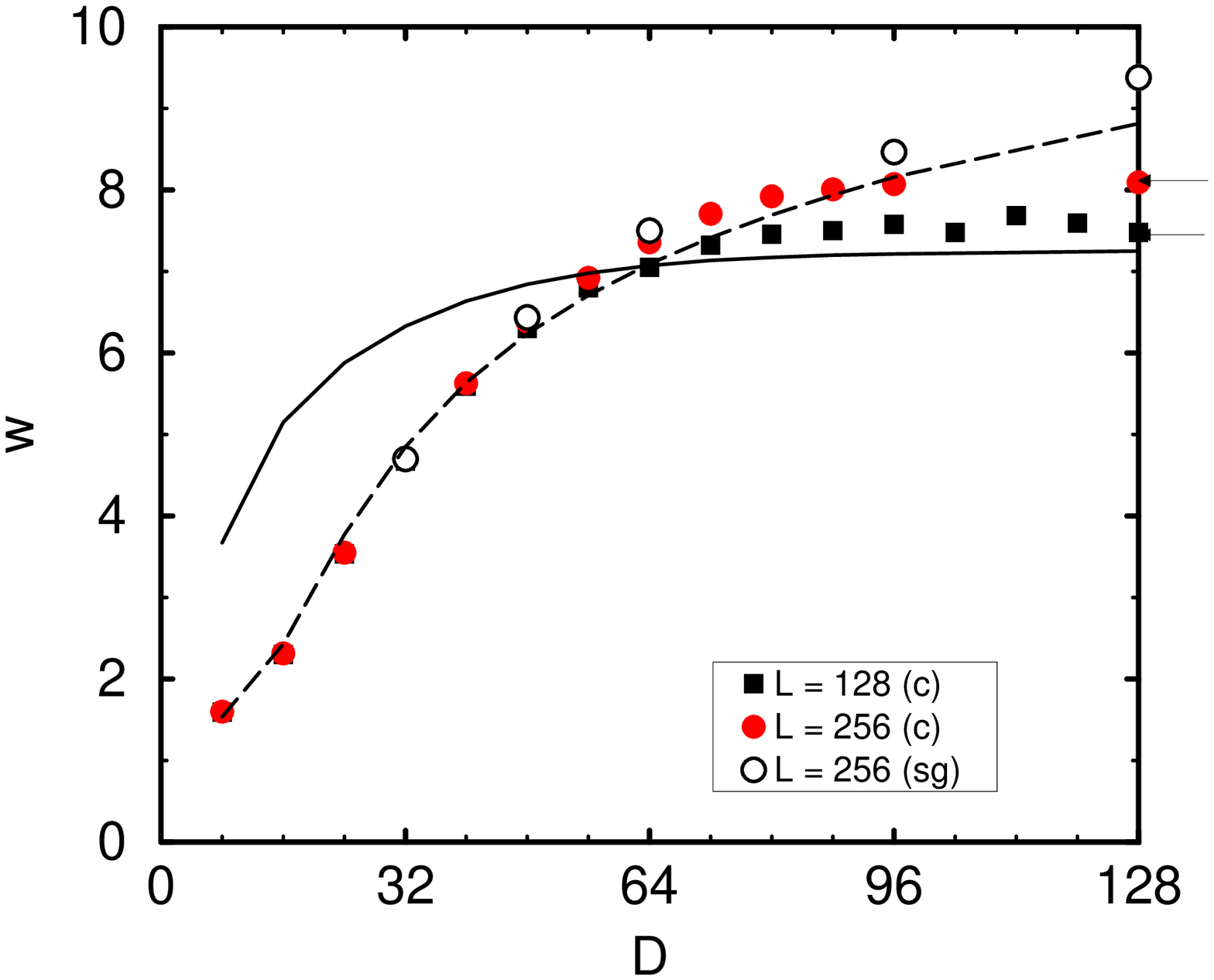}
           \setlength{\epsfxsize}{9cm}
           \epsffile{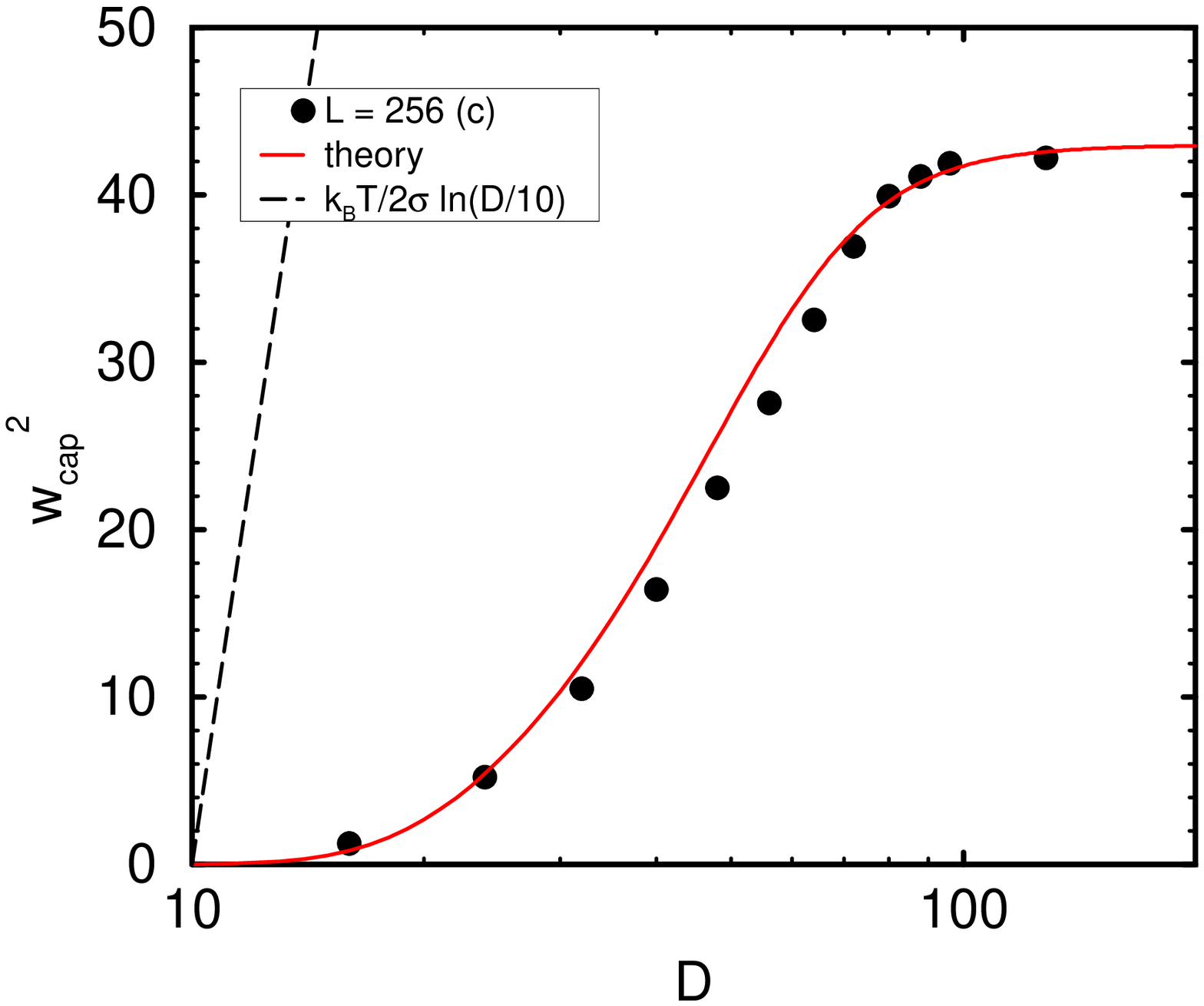}
       }
    \end{minipage}%
    \hfill%
    \begin{minipage}[b]{160mm}%
       \caption{
       \label{fig:width}
       }
		Dependence of the apparent interfacial width on the film thickness $D$ at $\lambda=5$.
                ({\bf a}) Filled symbols denote the apparent interfacial width in the canonical simulations, and the arrows on the right hand side mark the width of an unconfined interface at the lateral
		system size $L$. Open circles display the results of the semi-grandcanonical simulations. The lines correspond to fits without considering the the modification of the intrinsic interfacial
		properties.
		({\bf b}) Broadening $w^2_{\rm cap} = w^2(L=256) - w^2(B=8)$ of the interfacial width upon increasing the film thickness. The symbols correspond 
		to the same Monte Carlo results as in ({\bf a}), whereas the solid curve presents a detailed description according to eqns.(\ref{eqn:W}) and (\ref{eqn:XC}). The dashed line corresponds to 
		a simple logarithmic dependence.
    \end{minipage}%
\end{figure}

Having determined the influence of the surface fields on the intrinsic interfacial profiles, we turn to the apparent interfacial width $w$ which is accessible in experiments.\cite{KRAMER,JONESEX,TK1}
Fig.\ref{fig:width}({\bf a}) presents the dependence of the apparent interfacial width on the film thickness for $\lambda = 5$. Both data for canonical simulations and for semi-grandcanonical ones are
displayed. Similar to prior studies of short range surface fields,\cite{AW} the apparent interfacial width in the canonical ensemble saturates with growing film thickness at a value of the free 
interface broadened by
capillary waves. The limiting broadening is determined by the cut-off $q_{\rm min} = 2\pi/L$ rather than the lateral correlation length. In the semi-grandcanonical ensemble, the average 
composition fluctuates, hence the interface in a thick film of small lateral extension is only weakly bound to the center of the film and undergoes a diffusive motion. This leads to a linear increase of the 
apparent interfacial width with the film thickness.\cite{AW}

Similar to the experimental procedure, we have tried to estimate the intrinsic properties of the interface by describing the MC data for the apparent width {\em via}
eqn.(\ref{eqn:W}).
Using\cite{comment} $B_0=12$ and $\xi_\|$ according to eqn.(\ref{eqn:XL}), we fit the data
(solid line) and obtain for the intrinsic width $w_0 = 2.1$ and the interfacial tension $\sigma=0.022$. The curve does not describe the MC results well and the extracted values
for the intrinsic width and the interfacial tension differs from the SCF results  by a factor 0.5 and 1.5, respectively. In a second attempt we assume $\xi_\| = {\rm const} D^2$ and 
leave the proportionality
constant as a fit parameter. In the experimental situation this procedure corresponds to an unknown Hamaker constant. The dashed line displays the resulting fit which captures the behavior 
at small film thicknesses, but describes the thick film data less satisfactorily.  The
extracted parameters for the intrinsic width and the interfacial tension deviate from the SCF values by a factor 0.3 and 0.7, respectively. Moreover, the value of the parallel correlation length
is overestimated by more than an order of magnitude. Therefore we conclude that our MC results cannot be described appropriately by a simple effective interfacial Hamiltonian and it is not possible 
to extract reliable 
values for the intrinsic interfacial profiles from such an analysis. This result emphasizes the importance of the modification of the intrinsic interfacial properties by the presence of the long range surface 
fields. 

However, using the measured values for the intrinsic interfacial width $w(B_0=8,D)$ extracted {\em via} the block analysis, the effective interfacial tension $\sigma_{\rm eff}(D)$ as 
determined via the capillary fluctuation spectrum, and the parallel correlation length $\xi_\|(D)$ according to eqn.(\ref{eqn:XC}) we can describe the MC results quantitatively with
the theoretical prediction (\ref{eqn:W}). This is displayed in Fig.\ref{fig:width}({\bf b}) where we present the increase of the interfacial width due to capillary waves 
$w^2_{\rm cap} \equiv w^2-w^2_0(D) = 1/8\sigma_{\rm eff}(D)\;\ln([2 \pi/B_0]^2+\xi_\|^{-2})/([2\pi/L]^2+\xi_\|^{-2})$.
The long dashed line marks the simple logarithmic dependence $(1/2\sigma)\ln D$ for comparison.

\section{Discussion and outlook.}
\noindent
We have presented detailed Monte Carlo simulations targeted to investigating the effect of long range surface fields on the interfacial profiles in thin 
binary polymer films. The effect of capillary waves has been demonstrated conclusively {\em via} the dependence of the interfacial width of the apparent profiles on the 
lateral coarse graining size $B$ and a measurement of the Fourier spectrum of interfacial fluctuations. Both the strength of the surface fields $\lambda$
and the thickness of the film $2.3 R_g \leq D \leq 18.5 R_g$ influence the apparent interfacial width. The surface fields give rise to a lateral correlation 
length $\xi_\|$ which acts as a long wavelength cut-off for the spectrum of capillary waves. This parallel correlation length incorporates two contributions --
a short range one which stems from the distortion of the interfacial profile near the wall and a part caused by the long range surface field. For large film thicknesses the long range
contribution dominates and we find $\xi_\| \sim D^2$. The short range distortion contributes notably only for small surface fields and film thickness ($D \approx 5 R_g$). This is confirmed by 
two independent measurements in the Monte Carlo simulation and a quantitative comparison to self-consistent field calculations.

However, our Monte Carlo simulations are not compatible with a simple logarithmic dependence of the width on the film thickness, which has been assumed in experiments by Sferrazza {\em et al.}\cite{JONESEX}
Instead, the Monte Carlo simulations show an additional influence of the surface fields on the intrinsic interfacial properties: 
upon decreasing the film width to $2.3 R_g$, the intrinsic interfacial profile becomes 
narrower and the interfacial tension increases by more than a factor of 2. An effect of the long range surface fields on the intrinsic properties persists up to film thicknesses which exceed the
radius of gyration by roughly a factor of 10.
Similar effects have been observed for short range interactions.\cite{AW,MB}
The modification of the intrinsic interfacial properties is confirmed 
by self-consistent field calculations, which yield reasonable quantitative agreement without any adjustable parameter. The corrections to the effective interfacial Hamiltonian are too 
strong in our MC simulations to extract the intrinsic interfacial properties from a simple logarithmic growth of the apparent interfacial width with the film thickness. Taking due 
account of the capillary wave broadening {\em and} the influence of the surface interactions on the intrinsic profile simultaneously we can describe our Monte Carlo results quantitatively.

This observation might also offer an explanation why the value of the interfacial tension extracted from the thickness dependence of thin film experiments by experiments of 
Sferrazza {\em et al.}\cite{JONESEX} is larger than estimated {\em via} mean field theory. Deviations from the behavior described by the effective interfacial Hamiltonian have also
been found in experiments of Kerle {\em et al.}\cite{TK1,TK2} In both experiments the simple theory overestimates the apparent interfacial width if the intrinsic ({\em i.e.}, self-consistent
field) results for the interfacial widths and tension are employed. Moreover, in these two experiments the range of film thickness are of the order of several $R_g$, a situation comparable to the 
systems studied here {\em via} Monte Carlo simulations. We believe that a  quantitative understanding of the interfacial properties in thin films calls for a joint analysis of the interplay 
between surface forces and capillary waves as well as of the influence of the surface forces on the intrinsic interfacial properties.

\subsection*{Acknowledgment}
\noindent
We have benefited from stimulating discussions with T. Kerle and J. Klein.
This work was supported by the DFG under grant Bi314-3, Bi314-17, by the Bundesministerium 
f\"ur Bildung, Wissenschaft, Forschung und Technologie (BMBF) under grant N$^0$ 03N8008C,
and the Graduiertenkolleg Supramolekularer Systeme.
Generous access to the CRAY T3E at the HLRZ, J\"ulich, and the HLR, Stuttgart, 
and the computer facilities of the ZDV, Mainz, and the RHR, Kaiserslautern, is 
gratefully acknowledged.

\newpage
\pagestyle{empty}

\begin{thebibliography}{99}
\bibitem{comment}      We employ $B_0=12$ instead of $B_0=8$ to describe approximatively the effect of the different boundary conditions
		       in the block analysis.\cite{AW} While at the system size $q_{\rm min}=2\pi/L$ periodic boundary conditions apply,
		       free lateral boundary conditions apply on the block scale $2\pi/B$.
\end{thebibliography}
\end{document}